\documentstyle[sprocl,epsf]{article}

\def\be{\begin{equation}}
\def\ee{\end{equation}}
\def\bea{\begin{eqnarray}}
\def\eea{\end{eqnarray}}
\def\half{\textstyle{\frac{1}{2}}}

\def\third{{\textstyle{\frac{1}{3}}}}

\def\mintp{\int\!\frac{d^4 p}{(2\pi)^4}}
\def\mintpp{\int\!\frac{d^4 p'}{(2\pi)^4}}
\def\mintk{\int\!\frac{d^4 k}{(2\pi)^4}}

\begin{document}

\title{Baryon Structure in a Covariant Diquark--Quark Model$^*$}

\author{G. Hellstern,}

\author{R. B\"aurle, U. Z\"uckert, R. Alkofer, H. Reinhardt}
\address{Institute for Theoretical Physics, Auf der Morgenstelle 14,\\ 
72076 T\"ubingen, Germany}


\maketitle\abstracts{The 
baryon structure is investigated  in a covariant diquark-quark 
model. In this approach baryons emerge as relativistic bound states 
of a constituent quark and a $0^{+}$ or $1^{+}$ diquark.
After solving the  Bethe-Salpeter Equation for the scalar diquark quark 
system  
in ladder approximation we couple various external currents to the
constituents of the baryon to probe its internal structure.
The quark and the diquarks are assumed to be confined which 
is implemented by suitable choices for the propagators.
This leads to nontrivial vertex functions between the constituents
and the external current.
Baryonic matrix elements are then evaluated to extract observable 
formfactors.}
  
\section{Introduction}

The main goal of our work is to understand the baryon structure 
up  to a momentum transfer of $Q^2\approx 1-2$ GeV. 
This momentum regime is 
particulary interesting because of the interplay between hadronic 
degrees of freedom on the one side and their intrinsic quark structure
on the other side.
To describe hadrons as bound states of quarks nonperturbative 
methods of QCD are unavoidable, whereas for large $Q^2$ 
the perturbative results of QCD should be met.
So we are looking for an interpolating model to describe 
baryon structure  at the intermediate energy region.

From the experimental side there is also an interest in this kind
of calculations, because proposed and already started experiments
(e.g. at COSY, ELSA, TJNAF) 
explore the baryon structure to a very high precision in this
momentum regime.

This workshop is devoted to  diquarks as a 
tool to parametrize complicated or unknown structures. We will 
exploit this philosophy in the following. Since a fully relativistic
Faddeev equation  which determines the nucleon as bound state of 
three quarks (bound by some kind of gluon interaction) is almost 
impossible to solve
it is worthwhile to study the approximation where two quarks are
bound to a diquark and interact with the third quark through quark
exchange. Such a picture of a baryon has been derived by 
path integral techniques in the context of the Nambu--Jona-Lasinio
model \cite{Rei90} and in the Global Color \mbox{Model \cite{Cahi89}.}\\
\noindent
{\footnotesize $^*$ Supported by COSY under contract 41315266.}

\section{Bethe-Salpeter Equation for Baryons}
The basic assumption  of our model is  that 
a baryon is a bound state of a constituent quark and a scalar ($0^{+}$)
or axialvector ($1^{+}$) diquark which are bound due to quark 
exchange \cite{Rei90}. The corresponding Bethe-Salpeter equation (BSE)
for the nucleon  is then given by
\footnote{We use a  Euclidean space formulation
with $\{\gamma_\mu,\gamma_\nu\} = 2 \delta_{\mu \nu}, 
\gamma_\mu^{\dagger}= \gamma_\mu$, \\
$p \cdot q = \sum_{i=1}^{4} 
= p_i q_i$ and assume isospin symmetry.},
\bea
\Phi(P,p) = \mintk g(-k-p)\gamma_5 S(-k-p) \gamma_5 S(\half P +k)
D(\half P -k)\Phi(P,k) \nonumber\\*
+\mintk g(-k-p)\gamma^\mu S(-k-p) \gamma_5 S(\half P +k)D^{\mu \nu}
(\half P -k)\Phi^\nu (P,k) \nonumber\\*
\Phi^\mu(P,p) = \mintk g(-k-p)\gamma_5 S(-k-p)\gamma^\mu S(\half P +k)
D(\half P -k)\Phi(P,k) \nonumber\\*
+ \mintk g(-k-p)\gamma^\lambda 
S(-k-p)\gamma^\mu S(\half P +k)D^{\lambda \rho}
(\half P -k)\Phi^\rho(P,k) \nonumber\\*
\label{BSE}
\eea
where $\Phi(P,p)$ and $\Phi(P,p)^\mu$ are  the baryon (scalar and axial 
vector)
vertex functions with amputated quark 
and diquark legs. The quark propagator is denoted by $S(q)$ and the 
diquark propagator by $D(q)$ and $D^{\mu \nu}(q)$ respectively.
$g(q)$ describes a possible 
extension of the diquark quark vertices. 
A solution of the Bethe-Salpeter equation which assumes a static 
quark exchange can be found in ref. \cite{Buc92}. 
Although $1^{+}$ diquarks are partially included in our  
\mbox{calculation \cite{TBP},}
we will restrict ourselves in the following to scalar diquarks so that 
only the first line of equation (1) survives.

Because  $\Phi(P,p)$ describes a  spin $\half$ particle it is 
useful to write it as a product of an amplitude 
$\Psi(P,p)$, which depends on the total and the relative momentum
of the nucleon, and a  
Dirac spinor $u(P,S)$: \cite{Mey94} 
\be
\Phi(P,p) := \Psi(P,p)u(P,S).
\ee
After inserting  this ansatz in the Bethe-Salpeter Equation (\ref{BSE}),
we multiply with the adjoint spinor  $\bar u(P,S)$ 
from the right hand side and sum over the nucleon spins.
It is the immidiately seen, that only the projections  
\be
\chi(P,p):=\Psi(P,p)\Lambda^+ , \qquad
\Lambda^+=\frac{(-iP \!\cdot\! \gamma+M_B)}{2M_B}
\label{LAMBDA+}
\ee
to positive energy survive.
The projected amplitude $\chi(P,p)$ is then by construction 
an eigenfunction of $\Lambda^+$,  
\be
\chi(P,p)=\Psi(P,p) \Lambda^+ = \chi(P,p)\Lambda^+.
\label{projektion}
\ee
This leads to the possibility to decompose further the 
Dirac structure of $\chi$ : 
The decomposition
\be
\chi(P,p)=\half (1-i\frac{\gamma\!\cdot\! P}{M_B})S_1
+\half (-i \gamma \!\cdot\!p_T
-\frac{1}{2M_B}(\gamma\!\cdot\! p \gamma\!\cdot\! P
-\gamma\!\cdot\! P \gamma\!\cdot\! p))S_2,
\ee
is the only one compatible with equation (\ref{projektion}).
$p^{\mu}_T = p^{\mu} +P^{\mu} \frac{P\!\cdot\!p}{M^2_B}$
denotes the transversal relative momentum.
The still unknown functions $S_1(P^2,pP,p^2)$ and  $S_2(P^2,pP,p^2)$
are of course determined by the solution of the BSE.

In the rest frame of the bound state (where the actual calculations
are done) we obtain
\be
 \chi_{\rm r.f.}(P,p) =
      \left( \begin{array}{cc}
      {\bf 1} S_1(P,p) & {\bf 0} \\
      \mbox{\boldmath $\vec \sigma \vec p$} S_2(P,p)&
         {\bf 0} \end{array} \right).
\ee
It is now obvious that the first two columns of the 
$4 \times 4$ matrix 
correspond to spinors with positive energy, spin up (``large'' component) 
and spin down (``small'' component) respectively.
In the calculation of baryonic matrix elements we will restrict 
ourselves, in a first approximation, therefore to $S_1$.

Finally we expand the scalar functions $S_j(P,p), j=1,2$ in terms of 
Gegenbauer Polynomials \cite{Nieu96}
\be
S_j(P,p) = \sum_{k=0}^{\infty} i^k f_k^j(\sqrt{p^2}) C_k^1(
\frac{p \cdot P}{\sqrt{P^2}\sqrt{p^2}}),
\ee
and solve the Bethe-Salpeter Equation in the rest frame of the nucleon 
either by diagonalization or by iteration.
In this way, the bound state mass is determined by the eigenvalue
of the integral equation, while the expansion coefficients $f_k^j$
correspond to the eigenvectors.

\subsection{Parametrization of confined constituents}
Since neither quarks nor diquarks have  ever been 
observed as asymptotic
states in experiments, there is not much justification necessary
to include this feature of QCD in a calculation of baryon properties.

We implement confinement by using propagators which have no 
poles.
In case of the quark propagator
this has to be understood as an effective parametrization
of the behavior obtained in  Dyson-Schwinger studies of \mbox{QCD
\cite{Rob94}.} 
The quark self-energies which appear
in
\be
S(p)=-i\gamma\!\cdot\! p \sigma_v(p)+\sigma_s(p)
=[i\gamma\!\cdot\! p A(p) +B(P)]^{-1}
\ee
are then choosen as
\be
\sigma_v(p)=\frac{1-\exp(-(1+\frac{p^2}{m_q^2})}{p^2+m_q^2}
\quad {\rm and}\quad
\sigma_s(p)=m_q \sigma_v(p).
\ee
Obviously the exponentials remove the mass poles 
however, thereby in\-tro\-duc\-ing an essential singularity at $p^2=-\infty$.
\\
In ref.\cite{Ben96} it has been shown that confined diquarks can be obtained 
when one uses an appropriate irreducible quark-quark kernel (beyond
ladder approximation).
We simulate this by using
\be
D(p)=\frac{-i Z(p)}{p^2+m_d^2}
\ee
where 
\be
Z(p)=1-\exp(-(1+\frac{p^2}{m_d^2})).
\ee
Note that an elementary scalar diquark would correspond to $Z \equiv 1$.
With such a prescription of confined constituents we especially 
get rid of the unphysical quark-diquark thresholds which 
plague baryon calculations within the pure 
hadronized NJL-model \cite{Hel95}. 
\\
To estimate the effects of an intrinsic diquark structure we
did the calculation with 
a pointlike quark-diquark coupling 
\be
g(k)= g_0
\ee
as well as with a diquark formfactor (see also ref. \cite{Kus96})
\be
g(k)= g_0\frac{\Lambda^2}{k^2+\Lambda^2},
\ee
where
$\Lambda$ is of the order of $2m_q$ and which simulates the diquark 
structure in a crude way.
%
\begin{figure}[t]
\centerline{{\epsfxsize 6cm \epsfbox{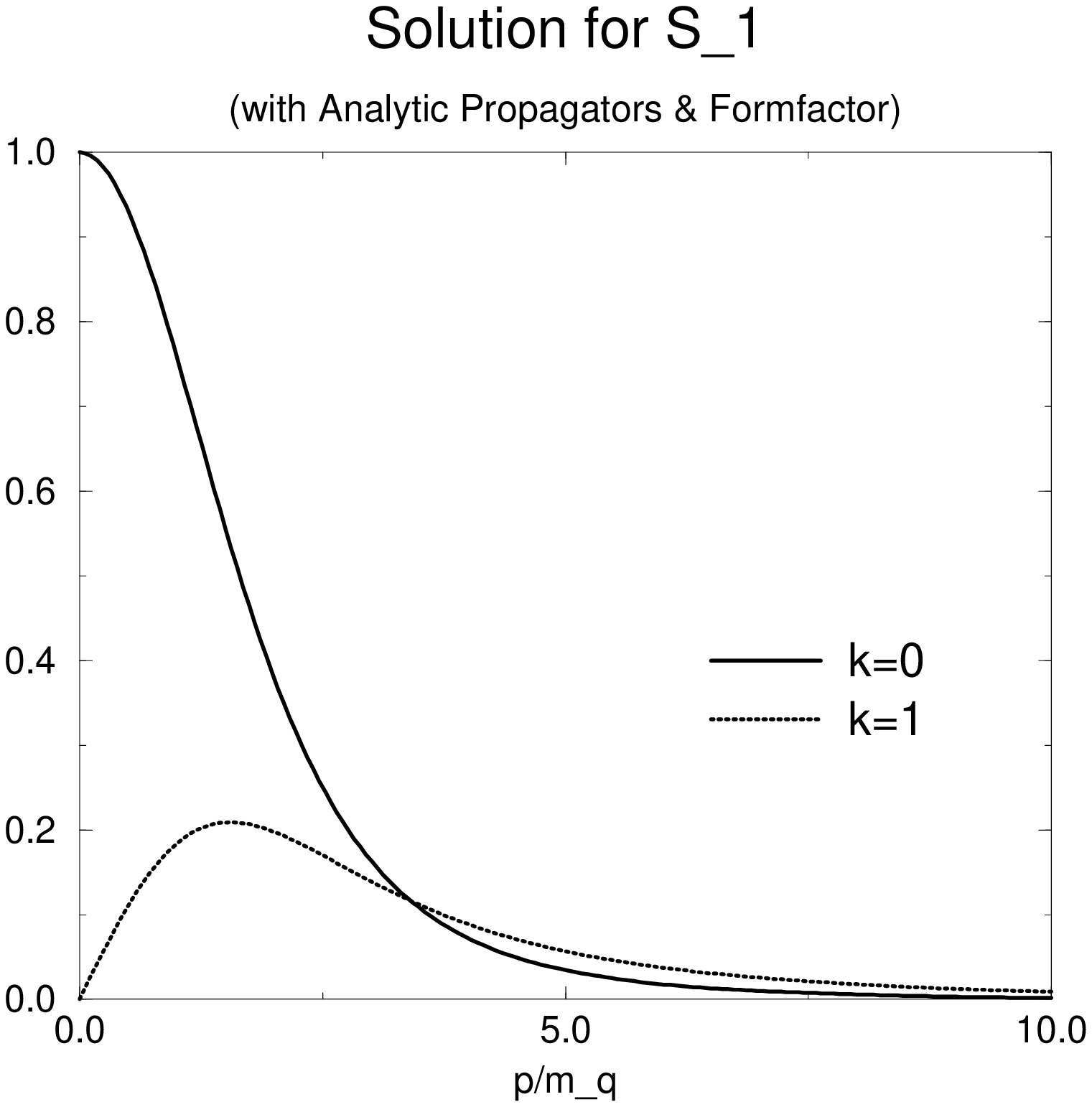}}{\epsfxsize 6cm
\epsfbox{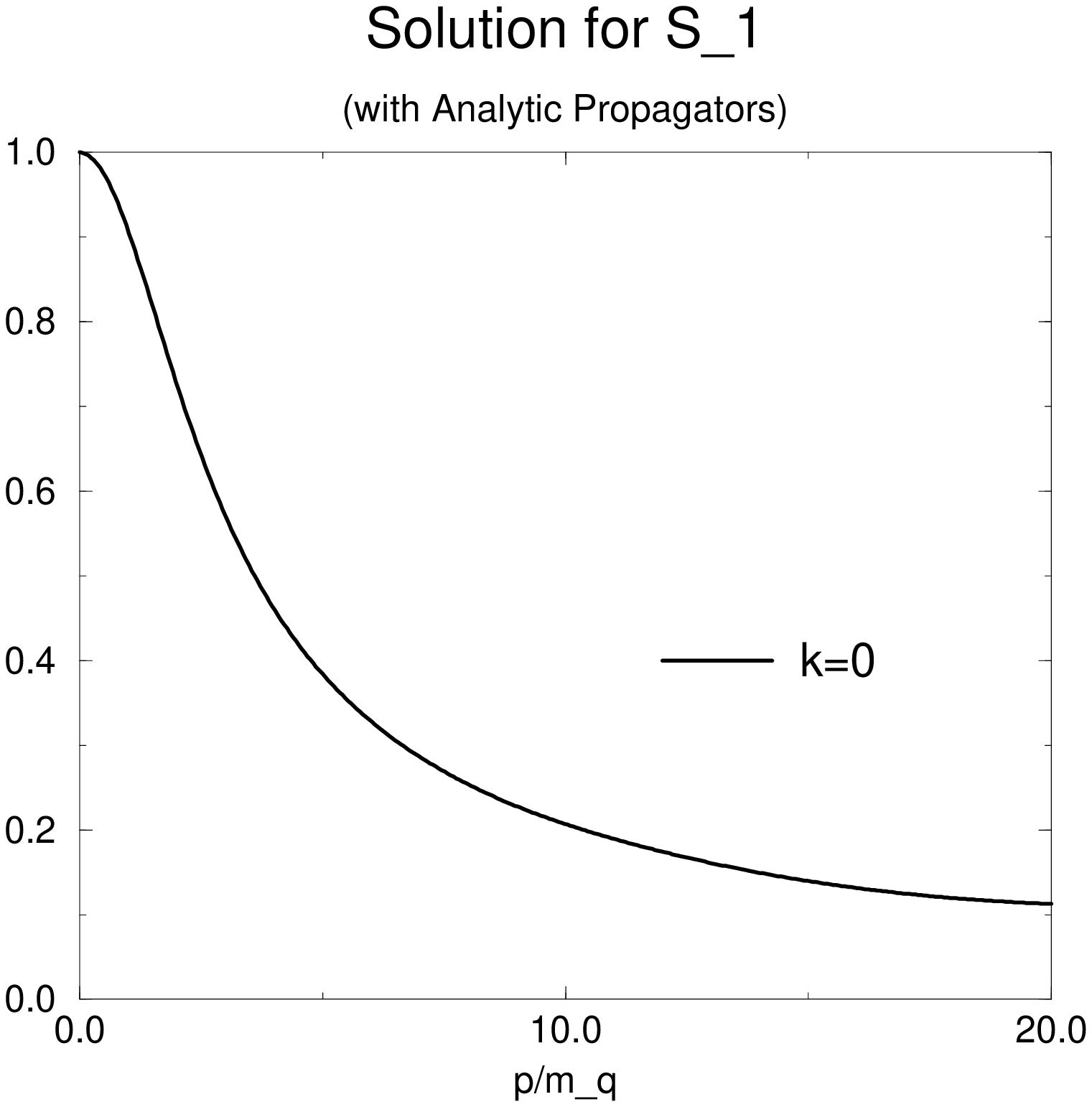}}} 
\caption{Numerical Solution of the Bethe-Salpeter equation.The results
in the left figure are obtained with a diquark formfactor, in the
right figure with a pointlike coupling.}
\label{fig1}
\end{figure}
%
\subsection{Numerical results}
In Fig.1  the numerical results for the dominant expansion functions 
$f_k^j$ without 
and with a diquark formfactor are shown. The parameters here 
and in the following  are fixed to $m_q=m_d = 0.5$ GeV
and the coupling  is chosen to obtain $M_B=0.95$ GeV.
In both figures the functions fall off rapidly for large relative
momenta, but the asymptotic behavior is quite different: Assuming a diquark
structure leads to a narrower amplitude in momentum space. 
This will also have an effect for the observables, where a 
possible diquark structure just enters  via the 
nucleon vertex
function.
We observe that in 
both cases the expansion in terms of Gegenbauer Polynomials 
converge rather rapidly.

\section{Hadronic matrix elements}
\subsection{Mandelstam's formalism}
Matrix elements of  operators in the Bethe-Salpeter approach  
are calculated according to Mandelstam's formalism \cite{Man55}
\be
\langle \hat O \rangle =  \mintp \mintpp
{\bar \psi} (P', p')\Gamma_{\hat O}(p',P';p,P) {\psi} (P,p).
\label{MANDELSTAM}
\ee
Note that $P$ and $p$ ($P'$ and $p'$) are the total and relative 
momenta of the diquark-quark states before and after the
interaction; ${\psi}$ is the 
normalized diquark-quark Bethe-Salpeter wave function related to the 
vertexfunction by multiplying with the quark and diquark propagator.
The 5-point function
$\Gamma_{\hat O}$ describes how the external current couples to all
internal lines of the diquark-quark system.
 
In the following we work in a generalized impulse approximation,
where we consider only the coupling to the quark and to the diquark, 
\be
\Gamma_{\hat O}(p',P';p,P)
=\Gamma^q_{\hat O}(p',P';p,P) + \Gamma^d_{\hat O}(p',P';p,P),
\ee
and neglect the coupling to the exchanged quark.

\subsection{Vertexfunctions of the constituents}
In case of electromagnetic (e.m.) formfactors, we still have to know
the  quark-photon and the diquark-photon vertexfunctions.
Using  gauge symmetry, which is ma\-ni\-fest in  the Ward-Takahashi identity,
\be
q_\mu i\Gamma_\mu(p,q,p+q) = S^{-1}(p+q)
-S^{-1}(p)
\label{WTI}
\ee
one observes the connection of the longitudinal part of the vertexfunction 
and the inverse propagator.
\\
A vertexfunction which solves the above identity for the quark is the 
Ball-Chiu vertex \cite{Bal80}
\bea
\Gamma_\mu^{BC}(p,k)=\frac{A(p^2)+A(k^2)}{2}\gamma_\mu
+\frac{(p+k)_\mu}{p^2-k^2} \{(A(p^2)-A(k^2))\frac{\gamma p + \gamma
k}{2} \nonumber\\*
-i(B(p^2)-B(k^2))\},
\label{BCV}
\eea
and in case of the diquark 
\be
\Gamma_\mu^{d}(p,k)=(p+k)_\mu \frac{D^{-1}(p)-D^{-1}(k)}{p^2-k^2}.
\ee
These vertexfunctions are totally determined by the self-energy
functions of the corresponding propagators.

\section{Electromagnetic nucleon formfactors}
%
\begin{figure}[t]
\centerline{{\epsfxsize 6.5cm \epsfbox{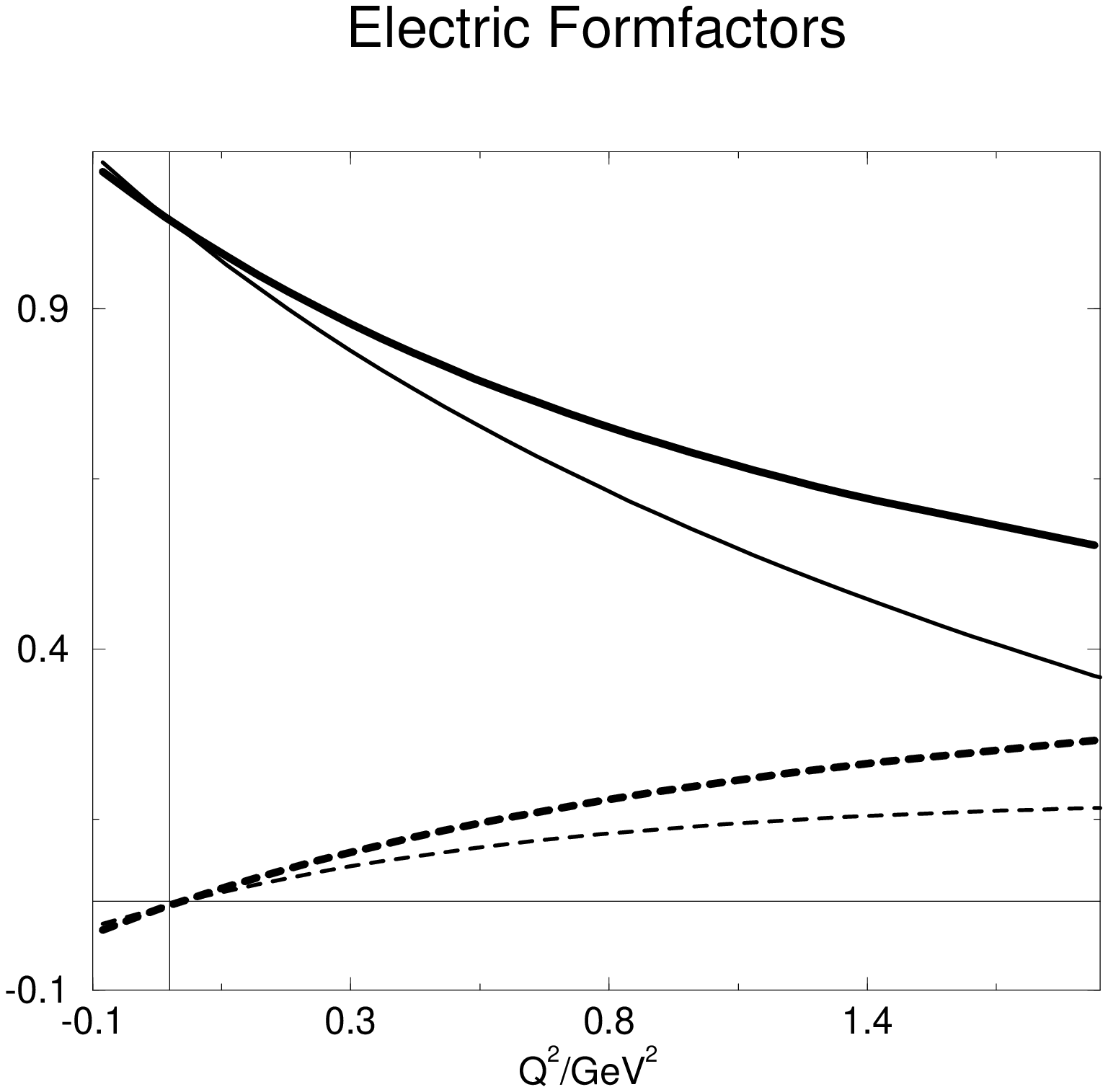}}{\epsfxsize 6.5cm
\epsfbox{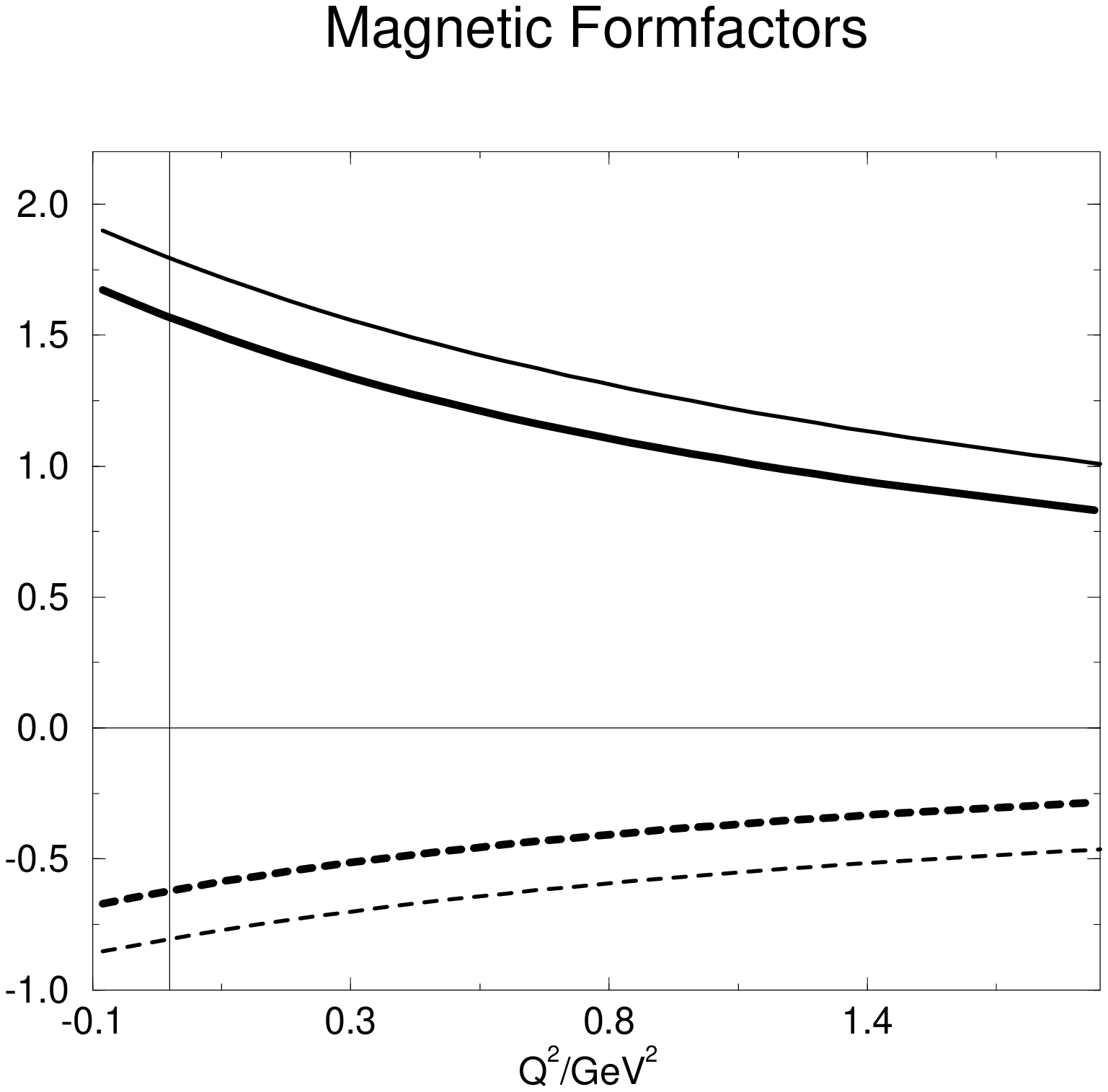}}} 
\caption{Electromagnetic Formfactors of the nucleon:  
The solid lines are the proton formfactors, the dotted lines
the neutron formfactors; the thin curves are the results with
a diquark formfactor, whereas the thick curves are the results
assuming a pointlike quark-diquark coupling.}
\label{fig2}
\end{figure}
%
Within our approximations we decompose the matrix element 
into a quark and a diquark part
\be
J_\mu(Q^2) = Q_q J_\mu^q(Q^2)+ Q_d J_\mu^d(Q^2),
\ee
with  the charge matrices in isospin space
\be
Q_q = \half(\third {\bf 1}_{\rm F} +\mbox{\boldmath $\tau$}_z)
,  \quad Q_d = \third {\bf 1}_{\rm F}.
\ee
The quark part is given by
\be
J_\mu^q(Q^2)=\mintk \bar \chi(P_f,p_f)S_q(k_+)i\Gamma_\mu^{BC}(k_+,
k_-)S_q(k_-)S_d(k_d)\chi(P_i,p_i),
\label{quarkpart}
\ee
whereas the diquark part is calculated with the loop diagram
\be
J_\mu^d(Q^2)=\mintk \bar \chi(P_f,p_f)S_d(k_+)i\Gamma_\mu^{d}(k_+,
k_-)S_d(k_-)S_q(k_q)\chi(P_i,p_i).
\label{diquarkpart}
\ee
From Lorentz invariance and discrete symmetries it is also 
known that the onshell e.m. nucleon current can be decomposed
into
\bea
J_\mu(Q^2) &=& \langle N(P_f,S_f)|j_\mu|N(P_i,S_i) \rangle
\nonumber\\*
&=&
\bar u(P_f,S_f)[iM_B(F_e(Q^2)-F_m(Q^2))\frac{P_\mu}{P^2}
+F_m(Q^2)\gamma_\mu]
u(P_i,S_i).\nonumber\\*
\eea
Note that in our formalism the Dirac spinors appearing on the right hand 
side are replaced by 
the $\Lambda^+$ projectors.
The Lorentz invariant functions 
$F_e(Q^2)$ and $F_m(Q^2)$
denote the electric and magnetic formfactor, respectively, which depend
on the photon momentum $Q^2$.
When the left hand side of this equation is calculated 
with the above loop diagrams 
in the Breit frame, $F_e(Q^2)$ and $F_m(Q^2)$ 
can be extracted by taking
appropriate traces.

In Fig.2 the numerical results for the e.m. formfactors
are shown. We observe that an internal diquark structure (leading
to a narrower vertex function in momentum space) seems 
to be necessary in our approach because otherwise the variation 
of the formfactors with $Q^2$ around $0$  is only  weak,
the e.m. radii are then far too smalll.
While the violation 
of charge conservation due to the impulse approximation
can be compensated by choosing a suitable  momentum distribution
in the loop integrals,
the absolute values of the magnetic moments are
not large enough
which means that $1^{+}$ diquarks are needed for a 
phenomenologically satisfying description.
Since our model is covariant, timelike formfactors are also accessible,
which can be seen in the figure: The formfactors are continous 
at $Q^2=0$.

\section{Pion-nucleon formfactor}
When following Mandelstam's prescripton it is also possible
to couple an external pion current to the nucleon. 
Because of parity there is no coupling to the (scalar) diquark,
so there is just one diagram to calculate (in analogy  to equation 
(\ref{quarkpart}) but where the quark-photon vertex is replaced by
the quark-pion vertex).
The quark-pion vertex ( = pion Bethe-Salpeter amplitude) in the chiral 
limit ($m_\pi=0$) is 
determined by the scalar quark selfenergy
\be
\Gamma_5^a(P^2=0,p^2)=\gamma_5 \frac{B(p^2)}{f_\pi}\tau^a,
\label{GAMMA5}
\ee
as a consequence of chiral symmetry \cite{Delb79}.
Furthermore  the invariant parametrization of the pion-nucleon
matrix element is given by 
\bea
J_5^a(Q^2) &=& \langle N(P_f,S_f)|j_5|N(P_i,S_i) \rangle
\nonumber\\
&=&
\bar u(P_f,S_f)[\gamma_5 g_{\pi NN}(Q^2)\tau^a]u(P_i,S_i).
\eea
In analogy to the electromagnetic formfactors, 
$g_{\pi NN}$ on the right hand 
side of this equation can be extracted when the left hand side 
is evaluated with the diagram where the pion couples to the quark
and the diquark remains a spectator.
\begin{figure}[t]
\centerline{{\epsfxsize 6.5cm \epsfbox{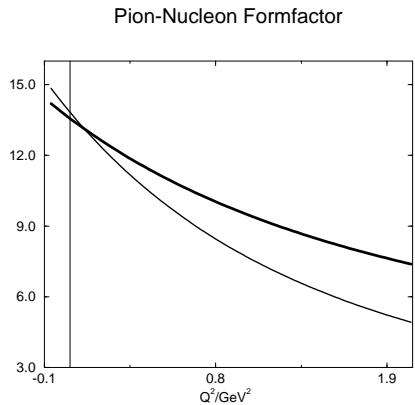}}} 
\caption{Numerical Solutions for the Pion-Nucleon coupling $g_{\pi NN}(Q^2)$.
The thin curve denotes our results with the diquark formfactor, the 
thick curve our results without a formfactor.}
\label{fig3}
\end{figure}
Our numerical results for $g_{\pi NN}$ are displayed in  Fig. 3.
Again a diquark structure is unavoidable 
to get the expected variation of the formfactor with the pion momentum. 
The absolute 
value of $g_{\pi NN}$ at the pion mass shell ($Q^2=0$ in the chiral limit)
is in remarkable 
agreement with the experimental value of about $14$.
\section{Conclusion}
In this talk we have presented a covariant bound state 
approach for the nucleon where quarks and diquarks interact through 
quark exchange. 
Using an effective parametrization of confinement we calculated 
within the needed approximations 
not only the nucleon mass and wave function but also 
electromagnetic and strong formfactors.
It has been discussed that an internal diquark structure, entering 
through a diquark formfactor in the BSE is preferable to obtain
a nucleon wave function which has a realistic width in momentum space.
Furthermore the implementation of axialvector diquarks is necessary
to improve the magnetic observables.

This work is just the first step towards a full satisfying 
baryon model of the proposed diquark-quark bound state approach. 
Nevertheless the results obtained so far are quite
encouraging.
To benefit from the advantages of our covariant and
confining approach we will also apply this  model to  
electromagnetic and strong $NN^*$ transitions,
virtual Compton scattering and $\Lambda$-production.
\section*{Acknowledgments}
G.H. would like to thank Prof.\ Anselmino and Prof.\ Predazzi 
for the very sti\-mu\-la\-ting and friendly athomosphere in Torino. 
He is also grateful to the Gra\-duier\-ten\-kolleg 
,,Hadronen und Kerne'' in T\"ubingen for 
financial support. 


\end{document}